\documentclass[a4paper, fleqn, useAMS, usenatbib]{mnras}
\usepackage[pdftex]{graphicx}
\usepackage{xcolor}
\usepackage{amsmath}
\usepackage{amsfonts}
\usepackage{amssymb}

\newenvironment{indentpar}[1]%
 {\begin{list}{}%
         {\setlength{\leftmargin}{#1}}%
         \item[]%
 }
 {\end{list}}

\def\q#1{`#1'}

\newcommand{\deaf}{`d', `e' and `f'}

\newcommand{\ap}{$\sim$ }
\title[YMCs and their Progenitors]{Comparing Young Massive Clusters and their Progenitor Clouds in the Milky Way}
\author[D. L. Walker et al.]{D.~L.~Walker$^{1, 2}$\thanks{E-mail: \texttt{D.L.Walker@2009.ljmu.ac.uk}}, 
S.~N.~Longmore$^{1}$,
N.~Bastian$^{1}$,
J.~M.~D.~Kruijssen$^{3,4}$,
\and J.~M.~Rathborne$^{5}$,
R.~Galv\'an-Madrid$^{6}$,
and H.~B.~Liu$^{7}$ \vspace{0.5cm}  \\
$^{1}$Astrophysics Research Institute, Liverpool John Moores University, IC2, 146 Brownlow Hill, Liverpool, L3 5RF, United Kingdom\\
$^{2}$Harvard-Smithsonian Center for Astrophysics, 60 Garden Street, Cambridge, MA 02138, USA \\
$^{3}$Max-Planck Institut fur Astrophysik, Karl-Schwarzschild-Strasse 1, 85748, Garching, Germany \\
$^{4}$Astronomisches Rechen-Institut, Zentrum f\"{u}r Astronomie der Universit\"{a}t Heidelberg, M\"{o}nchhofstra\ss e 12-14, 69120 Heidelberg, Germany\\
$^{5}$CSIRO Astronomy and Space Science, P.O. Box 76, Epping NSW, 1710, Australia\\
$^{6}$Instituto de Radioastronom\'\i a y Astrof\'\i sica, 
UNAM, Apdo. Postal 3-72 (Xangari), 58089 Morelia, Michoac\'an, M\'exico\\
$^{7}$3 Academia Sinica, Institute of Astronomy and Astrophysics, P.O. Box 23-141, Taipei 106, Taiwan\\
\\
}

\begin{document}

\date{}

\pagerange{\pageref{firstpage}--\pageref{lastpage}} \pubyear{2014}

\maketitle

\label{firstpage}
\begin{abstract}

Young massive clusters (YMCs) have central stellar mass surface densities exceeding 10$^{4}$ M$_{\odot}$ pc$^{-2}$. It is currently unknown whether the stars \emph{formed} at such high (proto)stellar densities. We compile a sample of gas clouds in the Galaxy which have sufficient gas mass within a radius of a few parsecs to form a YMC, and compare their radial gas mass distributions to the stellar mass distribution of Galactic YMCs. We find that the gas in the progenitor clouds is distributed differently than the stars in YMCs. The mass surface density profiles of the gas clouds are generally shallower than the stellar mass surface density profiles of the YMCs, which are characterised by prominent dense core regions with radii \ap 0.1~pc, followed by a power-law tail. On the scale of YMC core radii, we find that there are no known clouds with significantly more mass in their central regions when compared to Galactic YMCs. Additionally, we find that models in which stars form from very dense initial conditions require surface densities that are generally higher than those seen in the known candidate YMC progenitor clouds. Our results show that the quiescent, less evolved clouds contain less mass in their central regions than in the highly star-forming clouds. This suggests an evolutionary trend in which clouds continue to accumulate mass towards their centres after the onset of star formation. We conclude that a \emph{conveyor-belt} scenario for YMC formation is consistent with the current sample of Galactic YMCs and their progenitor clouds.

\end{abstract}


\section{Introduction}\

Young massive clusters (YMCs) are gravitationally bound stellar systems with masses $\gtrsim$ 10$^{4}$ M$_{\odot}$ and ages $\lesssim$ 100 Myr \citep{ymc_port}. Their masses and stellar densities can reach and even exceed those of globular clusters. Observations show that the cluster mass distribution is in fact continuous \citep{Larsen_CMF09, ymc_port}, extending from low-mass open clusters (\ap 100 M$_{\odot}$) to high-mass YMCs that are seen to be as massive as \ap 10$^{8}$ M$_{\odot}$ \citep[e.g. W3 in NGC 7252,][Cabrera-Ziri et al. 2016, in press]{W3}. This has potentially important implications, suggesting that clusters form in a similar way across this entire mass range. Additionally, it has been proposed that high-mass YMCs may be local analogues to the old globular clusters that we see today \citep[e.g.][]{elm_gcs,Diederik_submitted}. In this scenario, only the clusters formed in the early Universe that belonged to the high-mass end of the continuum would have been able to survive for a Hubble-time, whereas the lower mass clusters would have been disrupted \citep[e.g.][]{Vesperini_GC,Fall_GC,kruijssen15b}. If these scenarios are indeed true, this places YMCs in an important context -- by understanding their formation and evolution, it may be possible to gain an insight into the formation of clusters across the full mass range, including that of globular clusters.

The mechanism via which YMCs form is not yet entirely understood (see the review by \citealt{snl_ymc}). Much of the relevant discussion in the literature debates the initial distribution of the \emph{stars} in YMCs. There are two prominent theories on how stars are born in these clusters.

One scenario suggests that the stars form in a bound, centrally-condensed population in an extremely compact natal gas cloud. Feedback processes then remove the remaining gas, decreasing the global gravitational potential and causing the cluster to expand towards its final, un-embedded phase \citep[see e.g.][see the recent review by \citealt{B&K15}]{Lada_stellar_associations,cluster_mass_loss,nate_gas_removal,B&K_pop}. This is a {\it{monolithic}} formation scenario for YMCs.

The other scenario is one in which stars and sub-clusters form in accordance with the observed hierarchical structure of their natal gas clouds. Indeed, the interstellar medium is known to be hierarchical and sub-structured \citep[e.g.][]{ISM_structure, elme08, Diederik_cfe}. A heightened star formation efficiency (SFE) towards the densest peaks in the gas/dust leads to gas exhaustion on local scales, causing stellar dynamics to eventually dominate \citep{Diederik_cluster_dynamics,cluster_fragmentation,snl_ymc,Dale15}. The subsequent hierarchical merging of these stars and sub-clusters results in a centrally-concentrated, bound cluster \citep{ymc_merger_formation,Tiger14}. This is a {\it{hierarchical}} mode of YMC formation.

Note that the above scenarios can be confusing and can even co-exist. For example, \citet{B&K15} show that a cluster may form \emph{monolithically} from an initially \emph{hierarchical} distribution of stars, given an initially high density and prompt merging of sub-structure (\textless \ 1~Myr).

These \q{monolithic vs. hierarchical} discussions on YMC formation also do not adequately address the density evolution of both the \emph{gas and stars}. One main difference between these two scenarios is whether or not the stars are expanding or contracting after their immediate formation -- i.e. \emph{are the stars in YMCs born at initially higher or lower densities than their final gas-free distributions?} In order to address this, we need to study and compare the spatial distribution of the stars in YMCs with that of the gas in their gas-phase precursors. To date, there have been very few candidate YMC precursor gas clouds identified. In order to constrain possible formation mechanisms, it is essential that such clouds are found and studied in detail such that we can begin to understand the initial conditions of YMC formation. Recent efforts to survey the Galactic plane at far-infrared and (sub)millimetre wavelengths, where these dense, cold clouds emit brightly, have led to the identification of a growing sample of potential YMC precursor clouds.

The Central Molecular Zone \citep[CMZ; inner \ap 200 pc of the Galaxy,][]{CMZ} is known to host both the Arches and Quintuplet YMCs \citep[see e.g.][]{Figer_arches,Stolte_arches}, evidence that massive, bound stellar clusters do indeed form in this region of the Galaxy. The CMZ also contains a substantial reservoir of dense molecular gas, which hosts a population of massive and compact molecular clouds. Several of these clouds, notably G0.253$+$0.016 \citep[see e.g.][]{Lis_brick1,Lis_brick2,Brick, Brick_jens, Brick_jill, Brick_KJ,Jill_pdf_2014}, clouds \deaf \ \citep{Immer, Bricklets, Walker15} and the star-forming Sagittarius B2 complex \citep{Gaume, sgrb2}, are all thought to be likely YMC progenitors at different evolutionary stages. With the exception of Sagittarius B2, these clouds are devoid of any widespread star-formation, despite their gas volume densities being in excess of those in proposed star formation relations that suggest that stars should form efficiently above such densities \citep[see e.g.][]{Lada_2012_sfr,snl_sf}.

The Galactic disk hosts a larger known population of YMCs, such as the NGC 3603, Westerlund 1, Trumpler 14 and red super-giant (RSG) clusters \citep[][see the review by \citealt{ymc_port}]{NGC3603, Clark_west, Sana_tr14, Ben_RSG}, that reside over a range of Galactocentric radii. There are almost certainly more disk YMCs that have not yet been discovered, presumably more-so on the far-side of the Galactic bar, where it is very difficult to detect stellar clusters reliably due to dust extinction and crowding. A significant number of potential YMC precursor clouds have also been identified in the Galactic disk, such as the W49 and W51 star-forming complexes \citep{Ginsburg_clouds}. It is interesting to note that none of the YMC precursor clouds known in the disk are quiescent -- they are all forming stars at a high rate. The fact that there are four quiescent clouds at the Galactic centre with similar masses, radii and densities to the star-forming clouds in the disk is puzzling and suggests that perhaps there is something suppressing star formation at the Galactic centre and that star formation requires initially higher gas volume densities in this environment \citep{snl_ymc, Diederik_cmz_sf, Jill_pdf_2014}.

In our previous work, we compared quiescent YMC precursor clouds with (proto)YMCs at the Galactic centre, in an effort to assess the validity of the aforementioned cluster formation scenarios \citep{Walker15}. We found that the YMC progenitors there are not dense enough nor are they centrally-concentrated enough to form an Arches-like \citep[$M$ = 2x10$^{4}$ M$_{\odot}$, R$_{eff}$ = 0.4 pc;][]{Arches_esp} YMC without further dynamical evolution, despite them being the most massive and dense quiescent clouds found in the Galaxy. This result suggested that, at the Galactic centre, a monolithic mode of YMC formation is not viable given the present-day mass distribution within these clouds. We instead suggested that a hierarchical build up and merging of stellar mass is more likely. 

Given that the Galactic centre is an extreme environment -- with density, temperature, pressure, cosmic ray ionisation rate and magnetic field strength ranging from a factor of a few to orders of magnitude greater than in the Galactic disk \citep{Diederik_highz} -- it seems plausible that our previous result may be one that is specific to such an environment. Observing and characterising any environmental dependence of YMC formation is crucial if we are to develop a complete understanding of how they form and evolve as a function of environment, and may also have implications for the formation of {\it{all}} stellar clusters across the full mass range. Here, we therefore extend our study out into the Galactic disk, combining our previous sample with the known YMCs and their likely progenitors in the disk.

\ 

In this paper, we discuss revised potential formation scenarios for YMCs. Whilst the discussion on monolithic vs. hierarchical formation may describe the initial stellar distribution, it doesn't sufficiently explain the concurrent density evolution of both the gas and stars during the formation of a YMC. Instead, we propose the following three scenarios for YMC formation --

\begin{indentpar}{0.6cm}

\

\q{\emph{Conveyor-belt}}: Gas and stars have initial density distributions that are lower than that of an un-embedded YMC. Evolution is defined by the concurrent collapse of the molecular cloud and on-going star formation.

\

\q{\emph{In-situ}}: Gas is initially at a similar density as the final YMC stellar distribution. Stars can form at this density with little-to-no expansion or contraction.

\

\q{\emph{Popping}}: Gas is initially at a higher density than the un-embedded YMC. Once the stellar population has formed, the cluster expels its gas content and expands towards its final density distribution.
\end{indentpar}

Table 1 summarises the relevant general properties of the stellar and gas distributions in these different scenarios.
%___________
\begin{table*}
  \begin{tabular}{cccc}
    \hline
    	&	Before significant star	&	Majority of stars formed, still  & Un-embedded stellar cluster \\
    & formation ($t_{gas, initial}$)	& deeply embedded ($t_{*, initial}$) & ($t_{*, final}$) \\ \hline
	\q{\emph{Conveyor-belt}} & $R_{gas, initial} > R_{*, final}$ & $R_{*, initial} > R_{*, final}$ & $R_{*, final}$\\ 
    \emph{In-situ} & $R_{gas, initial} \approx R_{*, final}$ & $R_{*, initial} \approx R_{*, final}$ & $R_{*, final}$\\
    \q{\emph{Popping}} & $R_{gas, initial} < R_{*, final}$ & $R_{*, initial} < R_{*, final}$ & $R_{*, final}$\\ \hline \hline
  \end{tabular}
 \caption{Summary of global properties of the gas and stellar content at three distinct phases in our three proposed formation scenarios for YMCs. $t_{gas, initial}$ denotes the stage at which the YMC precursor cloud has not yet formed the majority of its stellar population. $R_{gas, initial}$ is the radius at this stage. The subscript (*, intial) refers to these properties at the stage at which most of the stars in the cluster have been formed, but the cluster is still embedded in the remaining gas. (*, final) indicates the final stage of YMC formation, where the cluster is completely free of gas.}
\end{table*}
%___________

\section{Data}\

For the Galactic YMC sample, we select all YMCs in \citet{ymc_port} with $M$ $\gtrsim$ 10$^{4}$ M$_{\odot}$ that have their surface density profiles published. Our sample of Galactic YMC precursor clouds is taken from those currently reported in the literature that satisfy the \citet{B12} criterion -- that the clouds have escape speeds larger than the sound speed in ionised gas. \citep{Ginsburg_clouds, atlasgal_hii, Bricklets}. 

\subsection{Galactic Centre}
\subsubsection{Clouds}\

Following from our previous work, we include the four quiescent Galactic centre clouds -- G0.253$+$0.016, \deaf. These clouds have recently been identified as potentially representing the early, starless phases of YMC precursors \citep{Brick, Bricklets, Brick_jill,Longmore16a}. We also include the gas/dust content surrounding the Sagittarius B2 Main and North proto-clusters. The analysis of these clouds is given in \citet{Walker15}. Data utilised are continuum data from the Herschel infrared Galactic Plane Survey \citep[HiGAL,][]{higal}, Bolocam Galactic Plane Survey \citep[BGPS,][]{bgps1,bgps2,bgps3} and ALMA project: ADS/JAO.ALMA\#2011.0.00217.S \citep{Jill_pdf_2014,rathborne15}.

\subsubsection{Clusters}\

The Galactic centre is known to host at least two YMCs (Arches and Quintuplet) and two possible proto-YMCs (Sagittarius B2 Main and North). We choose to implement a cluster age threshold at the Galactic centre of 2 Myr. This is chosen as the cluster disruption time-scale in this environment is very short, occurring over only a few to 10 Myr as a result of the strong tidal field and, most importantly, the disruptive tidal interactions with the dense gas \citep[e.g.][]{kim_gc_disrup, pz_gc_disruption,Diederik_cmz_sf}. Since we want to compare YMC progenitor clouds with the initial conditions of the stellar content of YMCs, we exclude the Quintuplet cluster from our sample, as it is already \ap 4 Myr old \citep{Figer_arches} and tidal disruption may have influenced the stellar surface density distribution.

The global properties and observed mass surface density profile for the Arches stellar cluster were obtained from \citet[][Table 8 and Figure 16]{Arches_esp}. Data for the stellar population of Sagittarius B2 proto-clusters (Main \& North) were extrapolated from the \citet{Gaume} radio observations of the embedded ultra-compact HII (UCHII) regions. See \citet{Walker15} for a detailed explanation of how these data were used to generate mass surface density profiles.

\subsection{Galactic Disk}
\subsubsection{Clouds}\

\citet{Ginsburg_clouds} and \citet{atlasgal_hii} report a sample of potential YMC precursor clouds in the Galactic disk. Of these, we select the clouds with $M$ $\gtrsim$ 3x10$^{4}$ M$_{\odot}$ within a radius of \textless \ 2.5 pc, as they could potentially form a 10$^{4}$ M$_{\odot}$ YMC with a global star formation efficiency of \ap 1/3. The clouds in our sample are G043.169$+$00.009 (W49), G049.489$+$00.386 (W51), G010.472$+$00.026, G350.111$+$0.089, G351.774$-$00.537 and G352.622$-$01.077.

As we are interested in the distribution of mass in these clouds and in particular whether they contain enough mass on small spatial scales to form a YMC at their present distributions, we require data with the highest possible angular resolution. For W49, we utilise high spatial-resolution (2") observations taken with the Submillimeter Array (SMA). The details of these observations and combination with single-dish data are discussed in \citet{W49_SMA}.

For both W51 and G010.472$+$00.026, we have extracted 450~$\mu$m SCUBA observations of these clouds from the JCMT data archive. These data provide a beam resolution of \ap 7". For the full details of these data, please refer to \citet{scuba}. We note that the effect of spatial filtering with SCUBA is not a concern here, as we are primarily interested in the gas and dust distribution on small spatial scales, where such filtering is not an issue.

For the remainder of the clouds in our sample, we extract (sub)mm continuum data for G350.111+0.089, G351.774-00.537 and G352.622-01.077 via the archive for the APEX Telescope Large Area Survey of the Galaxy \citep[ATLASGAL,][]{atlasgal_ref}. These data provide a beam resolution of \ap 19".

As we are using these data to generate mass surface density profiles for the clouds, we convert the data from units of flux to units of mass. We do this using the following relation (taken from \citealt{Jens_mass_in}, equation A.31, appendix A) --
\begin{equation}
\begin{split}
M = 0.12 M_{\odot}  \left( {\rm e}^{1.439 (\lambda / {\rm mm})^{-1}
      (T / {\rm 10 ~ K})^{-1}} - 1 \right) \\
   \cdot \left( \frac{\kappa_{\nu}}{0.01 \rm ~ cm^2 ~ g^{-1}} \right)^{-1} 
   \left( \frac{F_{\nu}}{\rm Jy} \right)
  \left( \frac{d}{\rm 100 ~ pc} \right)^2
  \left( \frac{\lambda}{\rm mm} \right)^{3},
\end{split}
\end{equation}

\

\noindent where $M$ is mass, $\lambda$ is wavelength, $T$ is the dust temperature, $k_{\nu}$ is the dust opacity, $F_{\nu}$ is the integrated flux and $d$ is distance (see Table 2 for assumed distances and respective references.). The dust temperature is not fully constrained observationally for these clouds. As such, we have to make certain assumptions regarding this parameter. Our assumption are as follows -- 

\begin{description}
  \item[(i)] The gas is isothermal throughout the extent of the cloud. This is a reasonable assumption for the Galactic centre clouds G0.253$+$0.016, \deaf, as they are quiescent. However the potential disk YMC precursors and Sagittarius B2 Main and North are all highly star-forming, and so we expect that there will be significant temperature gradients throughout the clouds.
  
  \item[(ii)] We assume that this single dust temperature is 20~K in all of the quiescent Galactic centre clouds. This is consistent with those measured from Herschel data \citep{Cara_higal, Walker15}. In all of the star forming clouds, we assume heightened dust temperatures of 40~K. The actual dust temperature will be much higher towards sites of star formation. Hence, any masses quoted are upper limits.
  
  \item[(iii)] We assume a constant gas-to-dust ratio of 100. We note that this may be lower by a factor of \ap 2 towards the Galactic centre \citep{snl_sf}.
\end{description}

The only remaining observationally unconstrained parameter in this relation is the dust opacity ($k_{\nu}$). To estimate this, we use the following relation, given in \S 3.2 of \citet{Cara_higal} -- 

\begin{equation}
k_{\nu} = 0.04 \rm ~ cm^{2} ~ g^{-1} \left(\frac{\nu}{505 ~ GHz}\right)^{1.75},
\end{equation}

\noindent where $\nu$ is the frequency. Note that this contains the explicit assumption that the gas-to-dust ratio is 100. \citet{Jens_mass_in} note that the uncertainties in both the dust temperature and opacity mean that the systematic uncertainties in mass estimates obtained via Equation 1 are \ap a factor of 2. See \citet{snl_sf} for a more in-depth discussion regarding the systematic uncertainties in obtaining mass estimates from dust emission.

Table 2 displays the general properties of the YMC precursor gas clouds in our sample. This shows that all of the clouds have similar global characteristics, with masses in the range of 10$^{4}$ -- 10$^{5}$ M$_\odot$, radii of \ap 2 -- 5 pc and volume densities of \ap 10$^{4}$ cm$^{-3}$.
 
\subsubsection{Clusters}\

We selected the YMCs given in Table 2 of \citet{ymc_port}, for which surface profiles are already published, specifically, NGC 3603, Westerlund 1 and Trumpler 14.

The data for NGC 3603 are taken from Figure 14 of \citet{Harayama_3603}. We take the data and fit for the surface density profile and take the assumed distance of 6.0 kpc to obtain their results in units of M$_\odot$ pc$^{-2}$. Given that their observations are sensitive to the mass range 0.5 -- 2.5 M$_{\odot}$, we extrapolate down to 0.1 M$_{\odot}$ and up to 120 M$_{\odot}$, assuming a Kroupa IMF, to obtain a corrective multiplicative factor of \ap 2.8. We then multiply the observed stellar mass surface density profile by this factor to retrieve the underlying total stellar mass surface density profile. We note that the effect of mass segregation has not been accounted for when applying IMF corrections to NGC 3603. However, as the observed mass range here is intermediate, we expect that the effect of mass segregation should be small.

Trumpler 14 data are taken from Figure 10 of \citet{Sana_tr14}. Their observations cover a mass range 0.1 -- 120 M$_{\odot}$ and so no IMF or mass segregation correction is necessary.

Westerlund 1 data are taken from Figure 8 of \citet{wd1}. The data are sampled from a mass range of 3.5 -- 32 M$_{\odot}$ -- extrapolating using a Kroupa IMF yields a multiplicative factor of \ap 4.4. We again multiplied the observed stellar mass surface density profile by this factor to retrieve the underlying total stellar mass surface density profile.

\section{Results}\

\begin{figure*}
\includegraphics[scale=0.82,angle=-90]{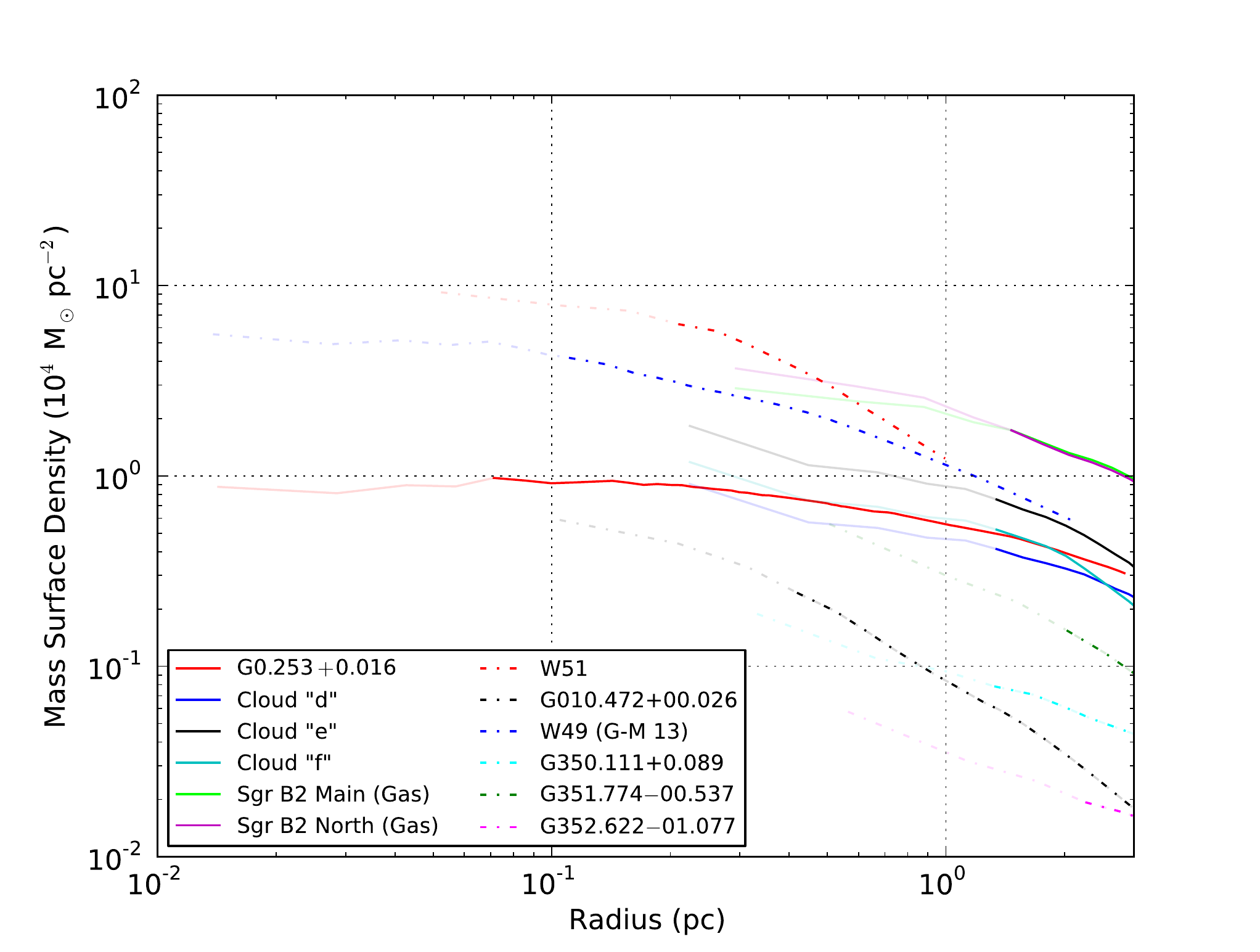} \\
 \caption{Enclosed mass surface density profiles as a function of radius for the YMC precursor clouds in the CMZ (solid lines) and in the Galactic disk (dashed lines). The transition point from solid to lower opacity indicates the beam resolution of the data.}
\end{figure*}

\begin{figure*}
\includegraphics[scale=0.82,angle=-90]{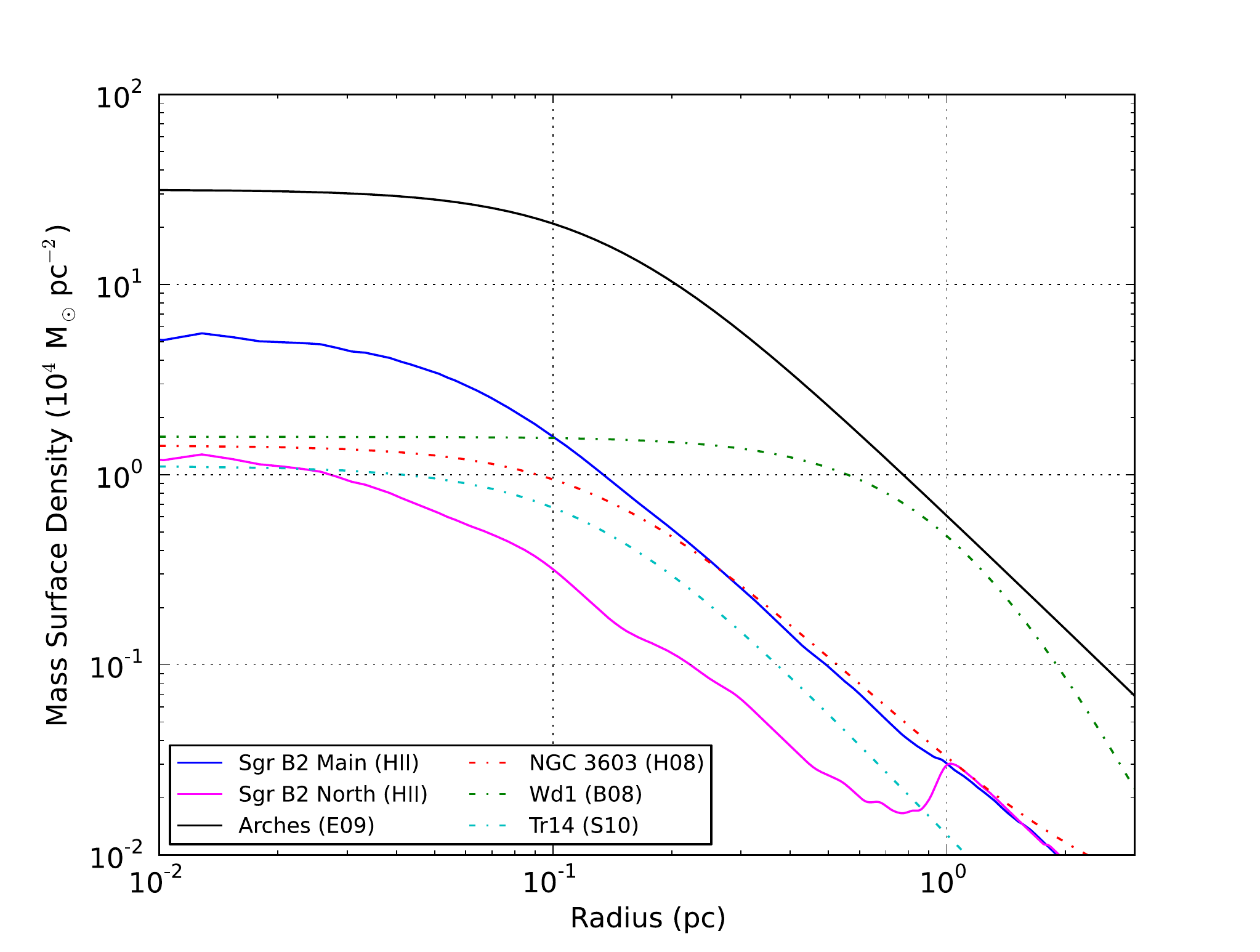} \\
 \caption{Enclosed mass surface density profiles as a function of radius for the Galactic YMCs along with the HII region distribution of the Sagittarius B2 Main (blue) and North (Magenta) proto-clusters. Arches data from \citet{Arches_esp} (E09; black). NGC 3603 data from \citet{Harayama_3603} (H08; red). Westerlund 1 data from \citet{wd1} (B08; green). Trumpler 14 data from \citet{Sana_tr14} (S10; cyan). Note that the bump in the profile for Sagittarius B2 North at R \ap 1~pc is due to Sagittarius B2 Main entering the aperture. Solid and dashed lines indicate that the clusters lie in the CMZ and Galactic disk, respectively.}
\end{figure*}

In Figure 1 we display the enclosed mass-surface-density profiles of the clouds. To obtain these profiles, we use the CASA software package \citep{casa} to take cumulative mass measurements within apertures of increasing radii, which are centred on the dust peaks, then dividing by the corresponding aperture area at each increment. As discussed in \citet{Walker15}, the Galactic centre clouds (solid lines) are similar in terms of their surface distributions -- flat (down to r \ap 0.2 - 0.3 pc), with no prominent central high-density region in the profile. ALMA data show that this continues down to \ap 0.015~pc for the cloud G0.253+0.016 \citep{Jill_pdf_2014}.  Observations using the Submillimeter Array (SMA) show a similar trend for clouds d, e and f (Walker et al., in prep). 

Figure 1 also shows the YMC progenitor candidate clouds in the Galactic disk (dashed lines). The enclosed mass surface density profiles of these clouds are similar to those of the Galactic centre clouds, in that they are relatively shallow across all spatial scales. We note that the profiles generated using the BGPS and APEX data may be affected by the angular resolution (33" and 19", respectively). However, we do not see any considerable differences in the shapes of the profiles when compared to those taken with higher resolution data.

%___________
\begin{table*}
\begin{center}
  \label{tab:global_properties}
  \begin{tabular}{cccccccccc}
    \hline
    Cloud	& M          & D   & R  & $n$ & N$_{H_2}$ & Reference\\
    & 10$^{4}$ M$_\odot$ & kpc & pc &10$^{4}$ cm$^{-3}$ & 10$^{24}$ cm$^{-2}$ & --\\ \hline
G0.253$+$0.016 &11.9   & 8.4$^{\dagger}$   & 2.9  & 1.7 & 0.6 & 1\\ 
    d  	&7.6     & 8.4$^{\dagger}$   & 3.2   & 0.8   & 0.3 & 1\\ 
    e 	&11.2      & 8.4$^{\dagger}$   & 2.4   & 2.8 & 0.9 & 1\\ 
    f   &7.3       & 8.4$^{\dagger}$   & 2.0   & 3.2 & 0.8 & 1\\ \hline
    W49 & 12.0 & 11.4 & 2.2 & 4.2 & 1.2 & 2\\
    W51 & 5.2 & 5.4 & 1.6 & 4.6 & 0.9 & 2\\ 
    G010.472+00.026 & 3.8 & 10.8 & 2.1 & 1.5 & 0.4 & 2\\
    G350.111+0.089 & 3.6 & 11.4 & 2.1 & 1.4 & 0.4 & 3\\
    G351.774$-$00.537 & 27 & 17.4 & 4.8 & 0.9 & 0.6 & 3\\
    G352.622$-$01.077 & 6.2 & 19.4 & 3.3 & 0.6 & 0.3 & 3\\
    \hline \hline
  \end{tabular}
  \caption{Global properties of the sample of likely YMC progenitor gas clouds used in this work. The columns show mass (M), distance (D), radius (R), average volume number density ($n$), average column density (N$_{H_2}$) and the corresponding reference. {\it{$^{\dagger}$Galactrocentric distance estimate from \citet{distance}  -- all Galactic centre clouds are assumed to be at this distance.}} References: [1] \citet{Walker15}, [2] \citet{Ginsburg_clouds} and [3] \citet{atlasgal_hii}.}
\end{center}
\end{table*}
%___________
\begin{table*}
\begin{center}
  \label{tab:global_properties}
  \begin{tabular}{cccccccc}
    \hline
    Cluster	& Age & M          & D   & R$_{core}$  & $\Sigma_{0}$ & Reference\\
    & Myr & 10$^{4}$ M$_\odot$ & kpc & pc &10$^{5}$ M$_\odot$ pc$^{-2}$ & --\\ \hline
Sagittarius B2 Main & \textless \ 0.5 & \textless \ 0.4   & 8.4   & \textless \ 0.1  & 0.5 & 1\\ 
Sagittarius B2 North & \textless \ 0.5 & \textless \ 0.4   & 8.4   & \textless \ 0.1  & 0.12 & 1\\ 
    Arches  & 2.0 	& 2.0     & 8.4 & 0.14   & 3.5   & 2\\ \hline
    NGC 3603 & 2.0 & 1.0 -- 1.6 & 6.0 & 0.14 & 0.15 & 3\\
    Trumpler 14 & \textgreater \ 0.3 & 0.4 -- 1.1 & 2.8 & 0.14 & 0.12 & 4, 5\\ 
    Westerlund 1 & 3.0 -- 5.0 & 2.0 -- 4.5 & 3.6 & 1.1 & 0.17 & 6\\
    \hline \hline
  \end{tabular}
  \caption{Global properties of the sample of Galactic YMCs used in this work. The columns show cluster age, mass (M), distance (D), core radius (R$_{core}$), central mass surface density ($\Sigma_{0}$)  and the corresponding reference. References: [1] \citet{Walker15}, [2] \citet{Arches_esp}, [3] \citet{Harayama_3603}, [4] \citet{Sana_tr14}, [5] \citet{A07_tr14} and [6] \citet{wd1}.}
\end{center}
\end{table*}
%___________

In analysing the mass distributions in these clouds, it is important to reiterate the assumptions that have gone in to estimating their masses. The assumption that the gas is isothermal is unlikely in the highly star-forming clouds in our sample. The energy injected into the gas via on-going star formation will cause these clouds to be internally heated. The effect of this will be to steepen their surface density profiles. The assumption of a single dust temperature is reasonable for the quiescent clouds in our sample, but less so for the star-forming clouds, where temperatures will be higher towards star-forming sites. Hence, any masses quoted are upper limits. We also assume a constant gas-to-dust ratio of 100, which may in fact be lower by a factor of \ap 2 towards the Galactic centre \citep{snl_sf}.

The result of our assumptions is that the steepness and mass-scaling of the profiles shown in Figure 1 are over-estimated -- they are strong upper limits. For example, if we assume that dust temperatures towards the star forming molecular cores are as high as 50 -- 200~K \citep[e.g.][]{core_temps}, this would result in a mass estimate that is \ap 2.5 -- 10 times lower towards these small regions. We note that the uncertainties in the W49 mass profile are different from the dust derived measurements, since this measurement comes from the ratios of CO isotopologues. \citet{W49_SMA} found that the effect of a radially decreasing gas temperature acts opposite to the effect of saturation in the innermost pixels (see their appendix D).

Table 3 shows the general properties of the YMCs in our sample. We see that they all have similar properties, in that they are all young ($\lesssim$ \ 2 Myr), \ap 10$^{4}$ M$_{\odot}$ and have core radii of order 0.1 pc. Westerlund 1 is the exception here (see \S 4 for discussion). These similarities are seen clearly in Figure 2, which displays their enclosed mass-surface-density profiles. Other than the differences in density and mass, all of the clusters (except Westerlund 1) have near-identical profile shapes, characterised by a prominent, compact central core region out to R \ap 0.1 pc.

Taking Figs. 1 and 2 and comparing them at face-value, it appears as though many of the clouds in our sample have equal or greater central mass surface densities when compared with the YMCs. This seems to suggest that perhaps some of  these clouds could form a YMC at their current densities. However, we note that our aforementioned assumptions regarding mass estimates place these as strong upper limits. We expect that both the central surface densities and the slope of the gas profiles will actually be lower. Furthermore, the profiles shown do not account for the fact that the star formation efficiency (SFE) must be smaller than 100\%. Even in high-density proto-stellar cores, the SFE is not expected to exceed 30\% -- 50\% \citep[e.g.][]{matzner00,alves07}. 

Comparing the \emph{shape} of the profiles displays a general difference in the way that mass is distributed in YMCs and the clouds. Except in the case of Westerlund 1, the stars in the YMCs follow a comparatively simple, spherical Plummer-like distribution \citep{Plummer}. They have very compact central regions that are surrounded by much larger envelopes, which display a clear power-law drop-off in surface density beyond the core scale (\ap 0.1~pc). In contrast, the molecular clouds have a much more uniform density over larger scales (i.e. no distinct central regions of high density). Their mass distribution is much more flat, with significant fall-off at radii approaching 1~pc. We note that this flat profile does not imply a lack of sub-structure on smaller spatial scales. High spatial-resolution ALMA continuum data reveal that clouds like G0.253$+$0.016 are highly sub-structured \citep{rathborne15}.

In Figure 3 we display the enclosed mass as a function of radius for all of the clouds in our sample. Solid and dashed lines correspond to Galactic centre and disk clouds, respectively. To compare the data to simulations of monolithic, or \q{popping} YMC formation, we also plot the initial conditions from several simulations in the literature. The triangle markers correspond to the initial conditions used by \citet{B&K1} to simulate the monolithic formation of the R136 YMC. The circular markers correspond to the initial conditions used by \citet{B&K14} to simulate the monolithic formation of the NGC 3603 YMC. The star markers correspond to the initial conditions given in the fifth row of Table 1 in \citet{Ass_popping}, in which they simulate the \q{popping} formation of massive clusters from very dense initial conditions. We choose this particular set of initial conditions as they are quoted to be sufficient to form a 10$^{4}$ M$_\odot$ cluster with a SFE \textless \ 20~\%.

As these simulations are 3D models, and the data that we discuss in this paper are 2D projections, we must take care to ensure that we are making a fair comparison to these models. In \citet{B&K1,B&K14}, they present their initial conditions as initial cluster mass, initial gas mass and half-mass radius. Using their assumed \ap 33\% star-formation efficiency, we combine these masses to obtain the total gass mass prior to star formation. We also use the following relation to convert the half-mass radius to the Plummer radius \citep{heggie&hut} --

\begin{equation}
R_{H}= \frac{R_{pl}}{\sqrt{2^{2/3}-1}},
\end{equation}

\

\citet{Ass_popping} already provide the Plummer radius. We then take the total 3D gas mass for all of the models and convert this to a projected mass \citep{heggie&hut} --
\begin{equation}
M(d)= M\left(1 + \frac{R_{pl}^2}{d^2}\right)^{-1},
\end{equation}

\

Overall Fig. 3 shows that the initial projected masses within a given radius for these various models generally too high, particularly at smaller radii. We note that the \citet{B&K14} model for NGC 3603 appears consistent with observations at radii \textgreater \ 0.5~pc. However, it is over-dense on smaller spatial scales. Whilst the models generally predict mass surface densities that are greater than the observed clouds, this is not unambiguously the case. The highly star-forming clouds such as Sagittarius B2, W49 and W51 all lie very close to these models. Though we reiterate that the mass estimates, particularly in the highly star-forming clouds, are upper limits. Accounting for temperature effects would create an even larger disparity between the observations and simulations. We also note that the simulated models assume that both the gas and stars are initially distributed according to a Plummer distribution, which is not true for the gas in the clouds in our sample (see \S 4). This is a critical difference between the true conditions in the ISM and those assumed in these models. The gravitational potential of the gas is dominant until a significant number of stars have formed within the gas, and it is clear that the initial potential does not arise from a Plummer-like gravitational potential.

\begin{figure*}
\includegraphics[scale=0.82,angle=-90]{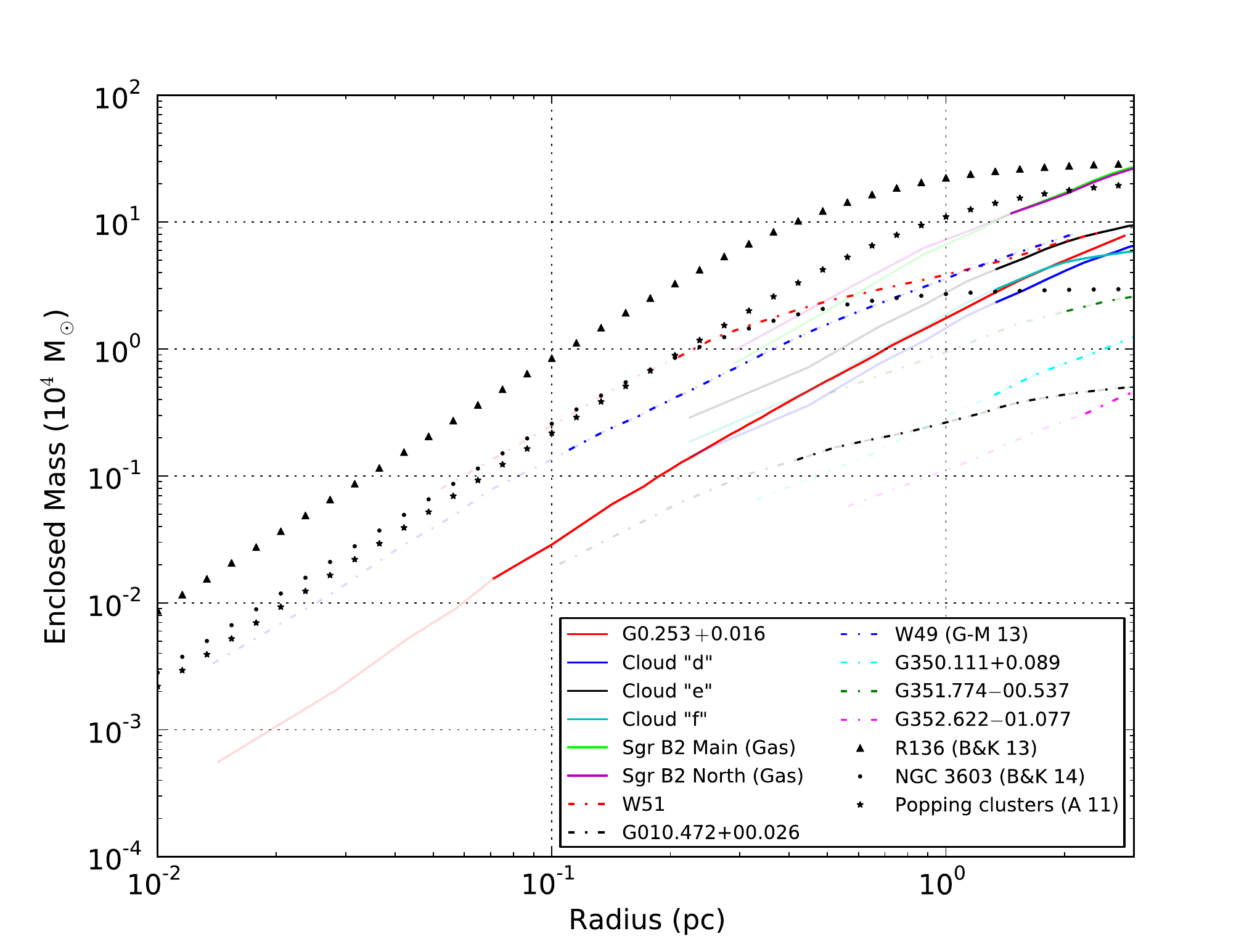} \\
 \caption{Enclosed gas mass profiles as a function of radius for the known potential Galactic YMCs precursors. The transition point from solid to lower opacity indicates the beam resolution of the data. The triangle markers correspond to the initial conditions used by \citet{B&K1} to simulate the monolithic formation of the R136 YMC. The circular markers correspond to the initial conditions used by \citet{B&K14} to simulate the monolithic formation of the NGC 3603 YMC. The star markers correspond to the initial conditions given in the fifth row of Table 1 in \citet{Ass_popping} to simulate a monolithically forming star cluster.}
\end{figure*}

\section{Discussion}\

In \citet{Walker15} we used a sample of YMCs and possible YMC precursor clouds at the Galactic centre to attempt to distinguish between two commonly argued modes of YMC formation -- {\it{popping}} or {\it{conveyor-belt}} modes. We did this simply by comparing their mass-surface-density profiles. Through comparing the surfaces profiles of the clouds and YMCs, it should be possible to begin to distinguish between these two potential modes of YMC formation. If {\it{popping}} cluster formation is a common mode, we should expect to see YMC precursor gas clouds that are highly centrally-concentrated, more-so than the YMCs, such that once formed they may expand out to the observed YMC stellar densities as a result of residual gas expulsion. If the {\it{conveyor-belt}} route is instead a viable mode of YMC formation, we should expect to see gas clouds that are more extended than YMCs and that show evidence for sub-structure on smaller scales. 

In this previous work, we found that in the Galactic centre, all of the candidate precursor clouds were more extended and less centrally-concentrated than the YMCs -- seemingly consistent with a {\it{conveyor-belt}} formation mode being more likely. Furthermore, sub-pc observations of one of these clouds, G0.253$+$0.016, reveal that it is indeed sub-structured on small spatial scales \citep{rathborne15}. Now that we have increased our sample size to include YMCs and potential YMC precursor clouds in the Galactic disk, we examine whether this conclusion holds true. 

%___________
\begin{table}
\begin{center}
  \label{tab:global_properties}
  \begin{tabular}{ccc}
    \hline
    Cloud	&	M$_{R=0.1}$	&	30\% SFE\\
    & 	10$^{3}$ M$_\odot$	&	10$^{3}$ M$_\odot$\\ \hline
	G0.253$+$0.016 & 0.29 & 0.09\\ 
    W49 & 1.28 & 0.43\\
    W51 & 2.71 & 0.90\\ 
    \hline
    Cluster & & \\
    \hline
    Arches & 6.6 & --\\
    NGC 3603 & 0.3 & --\\
    Trumpler 14 & 0.2 & --\\
    Westerlund 1 & 0.5 & --\\
    \hline \hline
  \end{tabular}
  \caption{Mass contained within a radius of 0.1~pc for the clouds (upper) and clusters (lower) in our sample. The right column shows this with an assumed 30\% global star formation efficiency (SFE) for the clouds.}
\end{center}
\end{table}
%___________

\subsection{\q{Popping} and in-situ clusters -- can Galactic clouds form a YMC at a high initial density?}

Using the clouds for which we possess high spatial-resolution data ($\lesssim$ 0.1~pc), we compare the mass contained within a radius of 0.1~pc -- the typical core radius of the YMCs in our sample -- in both the clouds and the YMCs. In doing this, we can assess whether these progenitor clouds contain enough mass on this scale to form a typical YMC stellar core at their present density distributions. The cluster core is by far the most dense region in these clusters, and so it follows that if they form in-situ, then the progenitors to such clusters should contain at least enough mass on the typical core scale such that they could form a stellar population that is at least as dense as the present-day populations in the central regions of Galactic YMCs.

The results are presented in Table 4. For the clouds, we also display this mass adjusted for an assumed upper limit of 30\% for the global star formation efficiency. We find that both W49 and W51 have comparable or greater central mass surface densities than NGC 3603, Trumpler 14 and Westerlund 1 at the typical core radius scale. This is also true for the highly star-forming Sagittarius B2 Main and North regions (Lu et al., in prep). In the largely-quiescent G0.253$+$0.016, we find that the cloud does not yet contain enough mass on this scale to form even the least centrally-dense YMC in our sample (Trumpler 14). A similar result is also found for the quiescent Galactic centre clouds ‘d’, ‘e’ and ‘f’, which are also quiescent (Walker et al., in prep).
 
These results suggest that in the \emph{evolved, star-forming} clouds, within \ap 10$^{5}$ years since the onset of star formation, sufficient mass has accumulated such that they could form a typical YMC stellar core in-situ. In contrast, the quiescent clouds have not yet had time to build up a dense enough mass reservoir in their central regions. We caution that in the Galactic disk, there is over an order of magnitude scatter in the central mass concentrations in the gas clouds. The same is true for the YMCs across all environments. This, coupled with the limited sample size of Galactic YMCs and their potential progenitors, as well as the lack of high-resolution observations towards many of the clouds, makes it difficult to infer any significant evolutionary trends. Nonetheless, we can say unequivocally that for the known sample there are no clouds that contain \emph{significantly} more mass than any known YMC in the central 0.1~pc. This is not compatible with a \q{\emph{popping}} formation scenario for YMCs, in which clusters form at initially higher densities, followed by a period of expansion due to gas expulsion. Instead, the apparent evolutionary trend from quiescent and less dense, to star-forming and more dense, suggests that we may be seeing evidence for a \q{\emph{conveyor-belt}} mode of YMC formation.
  
We also highlight that the core of the Arches cluster is considerably more dense than anything else in our sample. There are no known clouds that would be capable of forming such a core in-situ.

\subsection{On the clumpy sub-structure of molecular clouds}

As noted in \S3, the simulations of monolithic cluster formation by \citet{Ass_popping} and \citet{B&K1,B&K14} invoke Plummer-sphere initial morphologies for distribution of both the gas and the stars in the simulated clusters. Indeed, it is well established that the stars in YMCs are well described by Plummer models (or more generally, EFF-models; \citealt{EFF87}). Whether the stars were {\it{formed}} according to this distribution, as per the \q{\emph{popping}} or \q{\emph{in-situ}} scenarios, is  much less certain. Furthermore, the assumption that the gas follows a Plummer-like distribution is questionable, and inconsistent with what is typically seen in the interstellar medium (ISM), which is observed to have a hierarchical structure \citep[e.g.][]{ISM_structure}. Indeed, much of the data used in our sample for this paper shows this. The ALMA observations of G0.253$+$0.016 reveal that it is highly sub-structured on small spatial scales \citep[spatial resolution \ap 0.07~pc,][]{rathborne15} and not at all Plummer-like. The same holds true for clouds in the solar neighbourhood, which are highly filamentary \citep[e.g.][]{Barnard_1905}.\footnote{We thank Amy Stutz for pointing out this reference.}

Figure 4 shows the same mass surface density profile as given in Figure 1 for G0.253$+$0.016. Also plotted is a range of Plummer spheres, with core radii in the range of 0.5 -- 2.0~pc. It is clear that, no matter the core radius, a Plummer sphere does not represent the distribution of gas in this cloud. SMA observations (resolution \ap 0.15~pc) of clouds \deaf \ also reveal complex sub-structure (Walker et al. in prep.). SMA observations of W49 (resolution \ap 0.1~pc) show that this region is also complex, with hierarchical structures and filaments \citep{W49_SMA}. Given a spatially-varying star-formation efficiency, it is possible that a Plummer-like stellar distribution could form from gas with a different distribution. Nonetheless, the models for monolithic cluster formation use Plummer-like profiles for the gas, and this is not seen in the observations. 

For the remainder of the clouds in our sample, we currently do not possess data with high enough spatial resolution to comment on their small-scale structure. However, given that we know the structure of the aforementioned clouds, and that the structure of the ISM is well-established, we expect that they will follow a similarly complex, sub-structured distribution (i.e. not Plummer-like).

\begin{figure}
\includegraphics[scale=0.44,angle=-90]{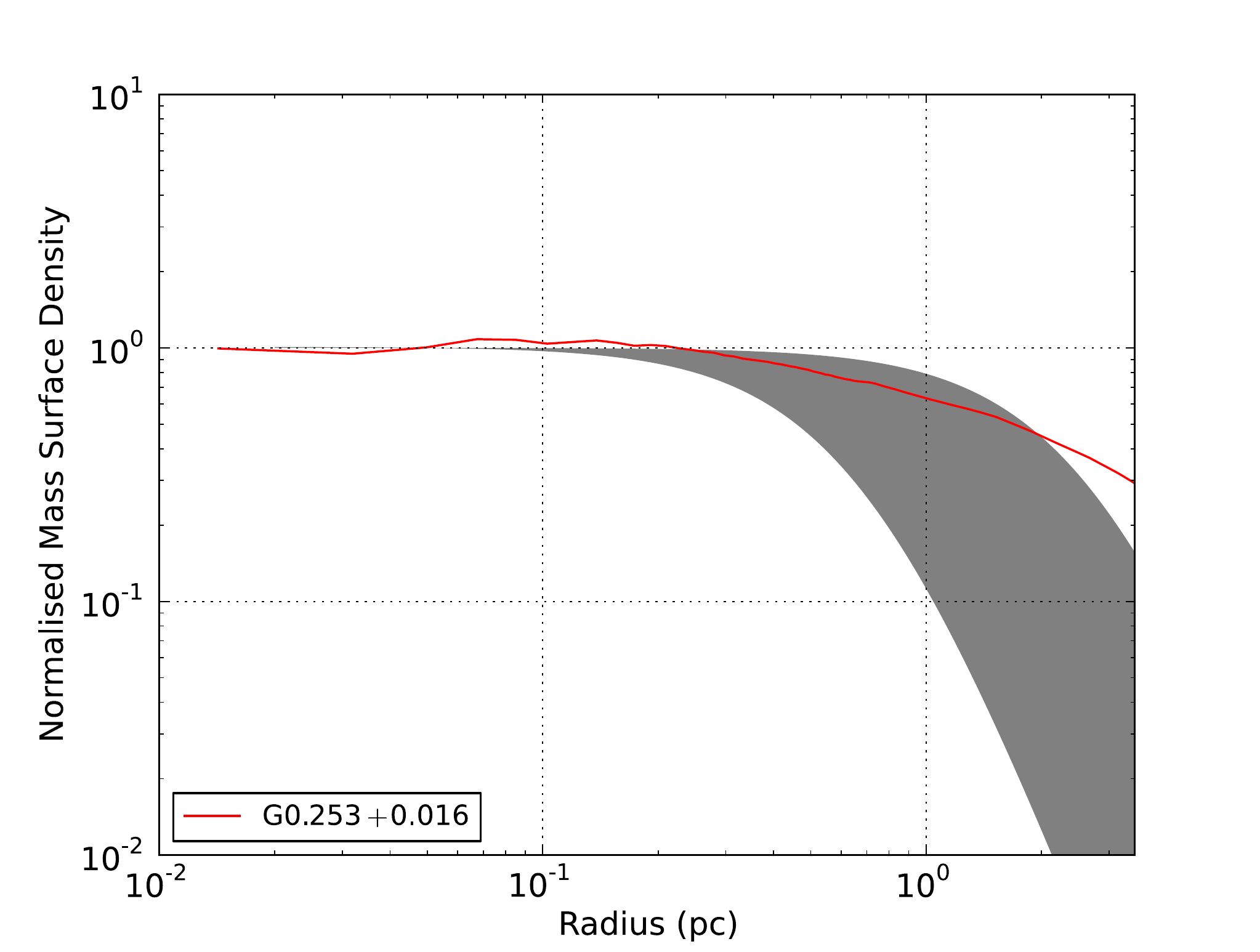} \\
 \caption{Mass surface density profile of G0.253$+$0.016. The shaded area corresponds to a range of Plummer spheres with core radii ranging from 0.5 -- 2~pc. We find that, no matter the core radius of the Plummer sphere, it cannot reproduce the mass surface density profile of the cloud -- the slope of the outer regions is too steep.}
\end{figure}

\subsection{The peculiar shape of Westerlund 1}

The stellar mass surface density profiles of the (proto)YMCs displayed in Figure 2 are all extremely similar in shape. They are well characterised by a Plummer-like profile with a core radius of \ap 0.1~pc. It is clear, however, that Westerlund 1 does not fit this description. The cluster is much more extended than the others in our sample, with a core radius of \ap 1.0 pc, almost an order of magnitude larger than the core radii of the other YMCs. This \q{bloated} appearance is further complicated by a reported elongation that is characterised by an axial ratio of 3:2 \citep{Wd1_elongation}. The source of this extended morphology is not known. However, \citet{Wd1_elongation} propose that it may be a result of merging of two or more stellar sub-clusters that formed in the same natal gas cloud. They reason that if this happened recently, the cluster may not yet have had time to dynamically relax and that eventually, this elongated cluster will settle into a more \q{normal} spherical distribution. If true, this may support the idea that a spherical, centrally-condensed cluster results through the merging of sub-clusters, as proposed in {\it{hierarchical merging}} scenarios for YMC formation. A corollary of this scenario is that all other YMCs in our sample have undergone violent relaxation. Irrespective of the core radius, all YMCs including Westerlund 1 display central surface densities similar to the proposed universal maximum surface density of $\Sigma_{*}\sim10^{5}~{\rm M}_\odot~{\rm pc}^{-2}$ observed in dense stellar systems over 7 orders of magnitude in stellar mass \citep{Hopkins_10}.

\section{Conclusions}\

We compare both the enclosed mass as a function of radius and the internal structure of the sample of known YMCs and their likely progenitor gas clouds throughout the Galaxy. We find that there are no known clouds with significantly more mass in their central regions than the known Galactic YMCs. The observations also show that the quiescent, less evolved clouds contain less mass in their central regions than the highly star-forming regions. This suggests an evolutionary trend in which clouds continue to accumulate mass towards their centres after the onset of star formation -- consistent with a \q{\emph{conveyor-belt}} mode of YMC formation.

When compared with simulations for monolithic \q{\emph{popping}} formation of YMCs, we find that the initial conditions for the cluster-forming clouds are not wholly consistent with the observations -- in general, they require more mass at a given radius than is observed in the known YMC precursor clouds. Furthermore, we find that the initial morphology of the gas in both these simulations and the general model for monolithic formation for YMCs is inconsistent with the observed morphology of YMC precursor clouds. They require initially Plummer-like, highly centrally-concentrated clouds -- whereas the clouds in our sample (for which high spatial resolution data are available) display  complex, hierarchical sub-structure and do not display the prominent cores and power-law tails of Plummer profiles. We therefore conclude that a \q{\emph{popping}} formation scenario for YMCs is not consistent with the data that is currently available for Galactic YMCs and their likely precursor gas clouds. We instead find that for the highly star-forming clouds, an \q{\emph{in-situ}} formation mode seems plausible. Coupled with the lower central densities of the quiescent clouds, this suggests a \q{\emph{conveyor-belt}}-like mode of YMC formation, whereby clouds contract and accumulate more mass in their central regions along with concurrent star formation.

\section*{Acknowledgements}

The authors would like to thank the referee, S\o ren Larsen, for his constructive comments, which were extremely helpful in clarifying and strengthening many aspects of this paper. This paper makes use of the following ALMA data: ADS/JAO.ALMA\#2011.0.00217.S. ALMA is a partnership of the European Southern Observatory (ESO) representing member states, Associated Universities Incorporated (AUI) and the National radio Astronomy Observatories (NRAO) for the National Science Foundation (NSF) in the USA, NINS  in Japan, NRC in Canada, and NSC and ASIAA in Taiwan, in cooperation with the Republic of Chile.  The Joint ALMA Observatory (JAO) is operated by ESO (Europe) , AUI/NRAO  (USA), and NAOJ (Japan). The National Radio Astronomy Observatory is a facility of the National Science Foundation operated under cooperative agreement by Associated Universities, Inc. This research made use of the NASA Astrophysical Data System.

\bibliography{bricklets_2}

\begin{thebibliography}{}
\makeatletter
\relax
\def\mn@urlcharsother{\let\do\@makeother \do\$\do\&\do\#\do\^\do\_\do\%\do\~}
\def\mn@doi{\begingroup\mn@urlcharsother \@ifnextchar [ {\mn@doi@}
  {\mn@doi@[]}}
\def\mn@doi@[#1]#2{\def\@tempa{#1}\ifx\@tempa\@empty \href
  {http://dx.doi.org/#2} {doi:#2}\else \href {http://dx.doi.org/#2} {#1}\fi
  \endgroup}
\def\mn@eprint#1#2{\mn@eprint@#1:#2::\@nil}
\def\mn@eprint@arXiv#1{\href {http://arxiv.org/abs/#1} {{\tt arXiv:#1}}}
\def\mn@eprint@dblp#1{\href {http://dblp.uni-trier.de/rec/bibtex/#1.xml}
  {dblp:#1}}
\def\mn@eprint@#1:#2:#3:#4\@nil{\def\@tempa {#1}\def\@tempb {#2}\def\@tempc
  {#3}\ifx \@tempc \@empty \let \@tempc \@tempb \let \@tempb \@tempa \fi \ifx
  \@tempb \@empty \def\@tempb {arXiv}\fi \@ifundefined
  {mn@eprint@\@tempb}{\@tempb:\@tempc}{\expandafter \expandafter \csname
  mn@eprint@\@tempb\endcsname \expandafter{\@tempc}}}

\bibitem[\protect\citeauthoryear{{Aguirre}, {Ginsburg}, {Dunham}  \& {et
  al}}{{Aguirre} et~al.}{2011}]{bgps2}
{Aguirre} J.~E.,  {Ginsburg} A.~G.,  {Dunham} M.~K.,   {et al} 2011, \mn@doi
  [\apjs] {10.1088/0067-0049/192/1/4}, \href
  {http://adsabs.harvard.edu/abs/2011ApJS..192....4A} {192, 4}

\bibitem[\protect\citeauthoryear{{Alves}, {Lombardi}  \& {Lada}}{{Alves}
  et~al.}{2007}]{alves07}
{Alves} J.,  {Lombardi} M.,   {Lada} C.~J.,  2007, \mn@doi [\aap]
  {10.1051/0004-6361:20066389}, \href
  {http://adsabs.harvard.edu/abs/2007A%26A...462L..17A} {462, L17}

\bibitem[\protect\citeauthoryear{{Ascenso}, {Alves}, {Vicente}  \&
  {Lago}}{{Ascenso} et~al.}{2007}]{A07_tr14}
{Ascenso} J.,  {Alves} J.,  {Vicente} S.,   {Lago} M.~T.~V.~T.,  2007, \mn@doi
  [\aap] {10.1051/0004-6361:20077210}, \href
  {http://adsabs.harvard.edu/abs/2007A%26A...476..199A} {476, 199}

\bibitem[\protect\citeauthoryear{{Assmann}, {Fellhauer}, {Kroupa}, {Br{\"u}ns}
  \& {Smith}}{{Assmann} et~al.}{2011}]{Ass_popping}
{Assmann} P.,  {Fellhauer} M.,  {Kroupa} P.,  {Br{\"u}ns} R.~C.,   {Smith} R.,
  2011, \mn@doi [\mnras] {10.1111/j.1365-2966.2011.18773.x}, \href
  {http://adsabs.harvard.edu/abs/2011MNRAS.415.1280A} {415, 1280}

\bibitem[\protect\citeauthoryear{{Banerjee} \& {Kroupa}}{{Banerjee} \&
  {Kroupa}}{2013}]{B&K1}
{Banerjee} S.,  {Kroupa} P.,  2013, \mn@doi [\apj]
  {10.1088/0004-637X/764/1/29}, \href
  {http://adsabs.harvard.edu/abs/2013ApJ...764...29B} {764, 29}

\bibitem[\protect\citeauthoryear{{Banerjee} \& {Kroupa}}{{Banerjee} \&
  {Kroupa}}{2014}]{B&K14}
{Banerjee} S.,  {Kroupa} P.,  2014, \mn@doi [\apj]
  {10.1088/0004-637X/787/2/158}, \href
  {http://adsabs.harvard.edu/abs/2014ApJ...787..158B} {787, 158}

\bibitem[\protect\citeauthoryear{{Banerjee} \& {Kroupa}}{{Banerjee} \&
  {Kroupa}}{2015}]{B&K15}
{Banerjee} S.,  {Kroupa} P.,  2015, \mn@doi [\mnras] {10.1093/mnras/stu2445},
  \href {http://adsabs.harvard.edu/abs/2015MNRAS.447..728B} {447, 728}

\bibitem[\protect\citeauthoryear{Barnard}{Barnard}{1905}]{Barnard_1905}
Barnard E.,  1905, PLATE 41, In Aquila, Northwest of Altair

\bibitem[\protect\citeauthoryear{{Bastian} \& {Goodwin}}{{Bastian} \&
  {Goodwin}}{2006}]{nate_gas_removal}
{Bastian} N.,  {Goodwin} S.~P.,  2006, \mn@doi [\mnras]
  {10.1111/j.1745-3933.2006.00162.x}, \href
  {http://adsabs.harvard.edu/abs/2006MNRAS.369L...9B} {369, L9}

\bibitem[\protect\citeauthoryear{{Battersby}, {Bally}, {Ginsburg}  \& {et
  al}}{{Battersby} et~al.}{2011}]{Cara_higal}
{Battersby} C.,  {Bally} J.,  {Ginsburg} A.,   {et al} 2011, \mn@doi [\aap]
  {10.1051/0004-6361/201116559}, \href
  {http://adsabs.harvard.edu/abs/2011A%26A...535A.128B} {535, A128}

\bibitem[\protect\citeauthoryear{{Baumgardt} \& {Kroupa}}{{Baumgardt} \&
  {Kroupa}}{2007}]{B&K_pop}
{Baumgardt} H.,  {Kroupa} P.,  2007, \mn@doi [\mnras]
  {10.1111/j.1365-2966.2007.12209.x}, \href
  {http://adsabs.harvard.edu/abs/2007MNRAS.380.1589B} {380, 1589}

\bibitem[\protect\citeauthoryear{{Boily} \& {Kroupa}}{{Boily} \&
  {Kroupa}}{2003}]{cluster_mass_loss}
{Boily} C.~M.,  {Kroupa} P.,  2003, \mn@doi [\mnras]
  {10.1046/j.1365-8711.2003.06101.x}, \href
  {http://adsabs.harvard.edu/abs/2003MNRAS.338..673B} {338, 673}

\bibitem[\protect\citeauthoryear{{Brandner}, {Clark}, {Stolte}, {Waters},
  {Negueruela}  \& {Goodwin}}{{Brandner} et~al.}{2008}]{wd1}
{Brandner} W.,  {Clark} J.~S.,  {Stolte} A.,  {Waters} R.,  {Negueruela} I.,
  {Goodwin} S.~P.,  2008, \mn@doi [\aap] {10.1051/0004-6361:20077579}, \href
  {http://adsabs.harvard.edu/abs/2008A%26A...478..137B} {478, 137}

\bibitem[\protect\citeauthoryear{{Bressert}, {Ginsburg}, {Bally}  \& {et
  al}}{{Bressert} et~al.}{2012}]{B12}
{Bressert} E.,  {Ginsburg} A.,  {Bally} J.,   {et al} 2012, \mn@doi [\apjl]
  {10.1088/2041-8205/758/2/L28}, \href
  {http://adsabs.harvard.edu/abs/2012ApJ...758L..28B} {758, L28}

\bibitem[\protect\citeauthoryear{{Cesaroni}, {Churchwell}, {Hofner}, {Walmsley}
   \& {Kurtz}}{{Cesaroni} et~al.}{1994}]{core_temps}
{Cesaroni} R.,  {Churchwell} E.,  {Hofner} P.,  {Walmsley} C.~M.,   {Kurtz} S.,
   1994, \aap, \href {http://adsabs.harvard.edu/abs/1994A%26A...288..903C}
  {288}

\bibitem[\protect\citeauthoryear{{Clark}, {Negueruela}, {Crowther}  \&
  {Goodwin}}{{Clark} et~al.}{2005}]{Clark_west}
{Clark} J.~S.,  {Negueruela} I.,  {Crowther} P.~A.,   {Goodwin} S.~P.,  2005,
  \mn@doi [\aap] {10.1051/0004-6361:20042413}, \href
  {http://adsabs.harvard.edu/abs/2005A%26A...434..949C} {434, 949}

\bibitem[\protect\citeauthoryear{{Dale}, {Ercolano}  \& {Bonnell}}{{Dale}
  et~al.}{2015}]{Dale15}
{Dale} J.~E.,  {Ercolano} B.,   {Bonnell} I.~A.,  2015, \mn@doi [\mnras]
  {10.1093/mnras/stv913}, \href
  {http://adsabs.harvard.edu/abs/2015MNRAS.451..987D} {451, 987}

\bibitem[\protect\citeauthoryear{{Davies}, {Figer}, {Kudritzki}  \& {et
  al}}{{Davies} et~al.}{2007}]{Ben_RSG}
{Davies} B.,  {Figer} D.~F.,  {Kudritzki} R.-P.,   {et al} 2007, \mn@doi [\apj]
  {10.1086/522224}, \href {http://adsabs.harvard.edu/abs/2007ApJ...671..781D}
  {671, 781}

\bibitem[\protect\citeauthoryear{{Di Francesco}, {Johnstone}, {Kirk}  \& {et
  al}}{{Di Francesco} et~al.}{2008}]{scuba}
{Di Francesco} J.,  {Johnstone} D.,  {Kirk} H.,   {et al} 2008, \mn@doi [\apjs]
  {10.1086/523645}, \href {http://adsabs.harvard.edu/abs/2008ApJS..175..277D}
  {175, 277}

\bibitem[\protect\citeauthoryear{{Elmegreen}}{{Elmegreen}}{2008}]{elme08}
{Elmegreen} B.~G.,  2008, \mn@doi [\apj] {10.1086/523791}, \href
  {http://adsabs.harvard.edu/abs/2008ApJ...672.1006E} {672, 1006}

\bibitem[\protect\citeauthoryear{{Elmegreen} \& {Efremov}}{{Elmegreen} \&
  {Efremov}}{1997}]{elm_gcs}
{Elmegreen} B.~G.,  {Efremov} Y.~N.,  1997, \mn@doi [\apj] {10.1086/303966},
  \href {http://adsabs.harvard.edu/abs/1997ApJ...480..235E} {480, 235}

\bibitem[\protect\citeauthoryear{{Elson}, {Fall}  \& {Freeman}}{{Elson}
  et~al.}{1987}]{EFF87}
{Elson} R.~A.~W.,  {Fall} S.~M.,   {Freeman} K.~C.,  1987, \mn@doi [\apj]
  {10.1086/165807}, \href {http://adsabs.harvard.edu/abs/1987ApJ...323...54E}
  {323, 54}

\bibitem[\protect\citeauthoryear{{Espinoza}, {Selman}  \& {Melnick}}{{Espinoza}
  et~al.}{2009}]{Arches_esp}
{Espinoza} P.,  {Selman} F.~J.,   {Melnick} J.,  2009, \mn@doi [\aap]
  {10.1051/0004-6361/20078597}, \href
  {http://adsabs.harvard.edu/abs/2009A%26A...501..563E} {501, 563}

\bibitem[\protect\citeauthoryear{{Fall} \& {Zhang}}{{Fall} \&
  {Zhang}}{2001}]{Fall_GC}
{Fall} S.~M.,  {Zhang} Q.,  2001, \mn@doi [\apj] {10.1086/323358}, \href
  {http://adsabs.harvard.edu/abs/2001ApJ...561..751F} {561, 751}

\bibitem[\protect\citeauthoryear{{Figer}, {Kim}, {Morris}, {Serabyn}, {Rich}
  \& {McLean}}{{Figer} et~al.}{1999}]{Figer_arches}
{Figer} D.~F.,  {Kim} S.~S.,  {Morris} M.,  {Serabyn} E.,  {Rich} R.~M.,
  {McLean} I.~S.,  1999, \mn@doi [\apj] {10.1086/307937}, \href
  {http://adsabs.harvard.edu/abs/1999ApJ...525..750F} {525, 750}

\bibitem[\protect\citeauthoryear{{Fujii}, {Saitoh}  \& {Portegies
  Zwart}}{{Fujii} et~al.}{2012}]{ymc_merger_formation}
{Fujii} M.~S.,  {Saitoh} T.~R.,   {Portegies Zwart} S.~F.,  2012, \mn@doi
  [\apj] {10.1088/0004-637X/753/1/85}, \href
  {http://adsabs.harvard.edu/abs/2012ApJ...753...85F} {753, 85}

\bibitem[\protect\citeauthoryear{{Galv{\'a}n-Madrid}, {Liu}, {Zhang}  \& {et
  al}}{{Galv{\'a}n-Madrid} et~al.}{2013}]{W49_SMA}
{Galv{\'a}n-Madrid} R.,  {Liu} H.~B.,  {Zhang} Z.-Y.,   {et al} 2013, \mn@doi
  [\apj] {10.1088/0004-637X/779/2/121}, \href
  {http://adsabs.harvard.edu/abs/2013ApJ...779..121G} {779, 121}

\bibitem[\protect\citeauthoryear{{Gaume}, {Claussen}, {de Pree}  \& {et
  al}}{{Gaume} et~al.}{1995}]{Gaume}
{Gaume} R.~A.,  {Claussen} M.~J.,  {de Pree} C.~G.,   {et al} 1995, \mn@doi
  [\apj] {10.1086/176087}, \href
  {http://adsabs.harvard.edu/abs/1995ApJ...449..663G} {449, 663}

\bibitem[\protect\citeauthoryear{{Gennaro}, {Brandner}, {Stolte}  \&
  {Henning}}{{Gennaro} et~al.}{2011}]{Wd1_elongation}
{Gennaro} M.,  {Brandner} W.,  {Stolte} A.,   {Henning} T.,  2011, \mn@doi
  [\mnras] {10.1111/j.1365-2966.2010.18068.x}, \href
  {http://adsabs.harvard.edu/abs/2011MNRAS.412.2469G} {412, 2469}

\bibitem[\protect\citeauthoryear{{Ginsburg}, {Bressert}, {Bally}  \&
  {Battersby}}{{Ginsburg} et~al.}{2012}]{Ginsburg_clouds}
{Ginsburg} A.,  {Bressert} E.,  {Bally} J.,   {Battersby} C.,  2012, \mn@doi
  [\apjl] {10.1088/2041-8205/758/2/L29}, \href
  {http://adsabs.harvard.edu/abs/2012ApJ...758L..29G} {758, L29}

\bibitem[\protect\citeauthoryear{{Ginsburg}, {Glenn}, {Rosolowsky}  \& {et
  al}}{{Ginsburg} et~al.}{2013}]{bgps3}
{Ginsburg} A.,  {Glenn} J.,  {Rosolowsky} E.,   {et al} 2013, \mn@doi [\apjs]
  {10.1088/0067-0049/208/2/14}, \href
  {http://adsabs.harvard.edu/abs/2013ApJS..208...14G} {208, 14}

\bibitem[\protect\citeauthoryear{{Girichidis}, {Federrath}, {Banerjee}  \&
  {Klessen}}{{Girichidis} et~al.}{2012}]{cluster_fragmentation}
{Girichidis} P.,  {Federrath} C.,  {Banerjee} R.,   {Klessen} R.~S.,  2012,
  \mn@doi [\mnras] {10.1111/j.1365-2966.2011.20073.x}, \href
  {http://adsabs.harvard.edu/abs/2012MNRAS.420..613G} {420, 613}

\bibitem[\protect\citeauthoryear{{Goss} \& {Radhakrishnan}}{{Goss} \&
  {Radhakrishnan}}{1969}]{NGC3603}
{Goss} W.~M.,  {Radhakrishnan} V.,  1969, \aplett, \href
  {http://adsabs.harvard.edu/abs/1969ApL.....4..199G} {4, 199}

\bibitem[\protect\citeauthoryear{{Harayama}, {Eisenhauer}  \&
  {Martins}}{{Harayama} et~al.}{2008}]{Harayama_3603}
{Harayama} Y.,  {Eisenhauer} F.,   {Martins} F.,  2008, \mn@doi [\apj]
  {10.1086/524650}, \href {http://adsabs.harvard.edu/abs/2008ApJ...675.1319H}
  {675, 1319}

\bibitem[\protect\citeauthoryear{{Heggie} \& {Hut}}{{Heggie} \&
  {Hut}}{2003}]{heggie&hut}
{Heggie} D.,  {Hut} P.,  2003, {The Gravitational Million-Body Problem: A
  Multidisciplinary Approach to Star Cluster Dynamics}

\bibitem[\protect\citeauthoryear{{Hopkins}, {Murray}, {Quataert}  \&
  {Thompson}}{{Hopkins} et~al.}{2010}]{Hopkins_10}
{Hopkins} P.~F.,  {Murray} N.,  {Quataert} E.,   {Thompson} T.~A.,  2010,
  \mn@doi [\mnras] {10.1111/j.1745-3933.2009.00777.x}, \href
  {http://adsabs.harvard.edu/abs/2010MNRAS.401L..19H} {401, L19}

\bibitem[\protect\citeauthoryear{{Immer}, {Menten}, {Schuller}  \&
  {Lis}}{{Immer} et~al.}{2012}]{Immer}
{Immer} K.,  {Menten} K.~M.,  {Schuller} F.,   {Lis} D.~C.,  2012, \mn@doi
  [\aap] {10.1051/0004-6361/201219182}, \href
  {http://adsabs.harvard.edu/abs/2012A%26A...548A.120I} {548, A120}

\bibitem[\protect\citeauthoryear{{Johnston}, {Beuther}, {Linz}, {Schmiedeke},
  {Ragan}  \& {Henning}}{{Johnston} et~al.}{2014}]{Brick_KJ}
{Johnston} K.~G.,  {Beuther} H.,  {Linz} H.,  {Schmiedeke} A.,  {Ragan} S.~E.,
   {Henning} T.,  2014, \mn@doi [\aap] {10.1051/0004-6361/201423943}, \href
  {http://adsabs.harvard.edu/abs/2014A%26A...568A..56J} {568, A56}

\bibitem[\protect\citeauthoryear{{Kauffmann}, {Bertoldi}, {Bourke}, {Evans}  \&
  {Lee}}{{Kauffmann} et~al.}{2008}]{Jens_mass_in}
{Kauffmann} J.,  {Bertoldi} F.,  {Bourke} T.~L.,  {Evans} II N.~J.,   {Lee}
  C.~W.,  2008, \mn@doi [\aap] {10.1051/0004-6361:200809481}, \href
  {http://adsabs.harvard.edu/abs/2008A%26A...487..993K} {487, 993}

\bibitem[\protect\citeauthoryear{{Kauffmann}, {Pillai}  \& {Zhang}}{{Kauffmann}
  et~al.}{2013}]{Brick_jens}
{Kauffmann} J.,  {Pillai} T.,   {Zhang} Q.,  2013, \mn@doi [\apjl]
  {10.1088/2041-8205/765/2/L35}, \href
  {http://adsabs.harvard.edu/abs/2013ApJ...765L..35K} {765, L35}

\bibitem[\protect\citeauthoryear{{Kim}, {Morris}  \& {Lee}}{{Kim}
  et~al.}{1999}]{kim_gc_disrup}
{Kim} S.~S.,  {Morris} M.,   {Lee} H.~M.,  1999, \mn@doi [\apj]
  {10.1086/307892}, \href {http://adsabs.harvard.edu/abs/1999ApJ...525..228K}
  {525, 228}

\bibitem[\protect\citeauthoryear{{Kruijssen}}{{Kruijssen}}{2012}]{Diederik_cfe}
{Kruijssen} J.~M.~D.,  2012, \mn@doi [\mnras]
  {10.1111/j.1365-2966.2012.21923.x}, \href
  {http://adsabs.harvard.edu/abs/2012MNRAS.426.3008K} {426, 3008}

\bibitem[\protect\citeauthoryear{{Kruijssen}}{{Kruijssen}}{2014}]{Diederik_submitted}
{Kruijssen} J.~M.~D.,  2014, \mn@doi [Classical and Quantum Gravity]
  {10.1088/0264-9381/31/24/244006}, \href
  {http://adsabs.harvard.edu/abs/2014CQGra..31x4006K} {31, 244006}

\bibitem[\protect\citeauthoryear{{Kruijssen}}{{Kruijssen}}{2015}]{kruijssen15b}
{Kruijssen} J.~M.~D.,  2015, \mn@doi [\mnras] {10.1093/mnras/stv2026}, \href
  {http://adsabs.harvard.edu/abs/2015MNRAS.454.1658K} {454, 1658}

\bibitem[\protect\citeauthoryear{{Kruijssen} \& {Longmore}}{{Kruijssen} \&
  {Longmore}}{2013}]{Diederik_highz}
{Kruijssen} J.~M.~D.,  {Longmore} S.~N.,  2013, \mn@doi [\mnras]
  {10.1093/mnras/stt1634}, \href
  {http://adsabs.harvard.edu/abs/2013MNRAS.435.2598K} {435, 2598}

\bibitem[\protect\citeauthoryear{{Kruijssen}, {Maschberger}, {Moeckel}  \& {et
  al}}{{Kruijssen} et~al.}{2012}]{Diederik_cluster_dynamics}
{Kruijssen} J.~M.~D.,  {Maschberger} T.,  {Moeckel} N.,   {et al} 2012, \mn@doi
  [\mnras] {10.1111/j.1365-2966.2011.19748.x}, \href
  {http://adsabs.harvard.edu/abs/2012MNRAS.419..841K} {419, 841}

\bibitem[\protect\citeauthoryear{{Kruijssen}, {Longmore}, {Elmegreen}  \& {et
  al}}{{Kruijssen} et~al.}{2014}]{Diederik_cmz_sf}
{Kruijssen} J.~M.~D.,  {Longmore} S.~N.,  {Elmegreen} B.~G.,   {et al} 2014,
  \mn@doi [\mnras] {10.1093/mnras/stu494}, \href
  {http://adsabs.harvard.edu/abs/2014MNRAS.440.3370K} {440, 3370}

\bibitem[\protect\citeauthoryear{{Lada}, {Margulis}  \& {Dearborn}}{{Lada}
  et~al.}{1984}]{Lada_stellar_associations}
{Lada} C.~J.,  {Margulis} M.,   {Dearborn} D.,  1984, \mn@doi [\apj]
  {10.1086/162485}, \href {http://adsabs.harvard.edu/abs/1984ApJ...285..141L}
  {285, 141}

\bibitem[\protect\citeauthoryear{{Lada}, {Forbrich}, {Lombardi}  \&
  {Alves}}{{Lada} et~al.}{2012}]{Lada_2012_sfr}
{Lada} C.~J.,  {Forbrich} J.,  {Lombardi} M.,   {Alves} J.~F.,  2012, \mn@doi
  [\apj] {10.1088/0004-637X/745/2/190}, \href
  {http://adsabs.harvard.edu/abs/2012ApJ...745..190L} {745, 190}

\bibitem[\protect\citeauthoryear{{Larsen}}{{Larsen}}{2009}]{Larsen_CMF09}
{Larsen} S.~S.,  2009, \mn@doi [\aap] {10.1051/0004-6361:200811212}, \href
  {http://adsabs.harvard.edu/abs/2009A%26A...494..539L} {494, 539}

\bibitem[\protect\citeauthoryear{{Larson}}{{Larson}}{1981}]{ISM_structure}
{Larson} R.~B.,  1981, \mnras, \href
  {http://adsabs.harvard.edu/abs/1981MNRAS.194..809L} {194, 809}

\bibitem[\protect\citeauthoryear{{Lis} \& {Menten}}{{Lis} \&
  {Menten}}{1998}]{Lis_brick2}
{Lis} D.~C.,  {Menten} K.~M.,  1998, \mn@doi [\apj] {10.1086/306366}, \href
  {http://adsabs.harvard.edu/abs/1998ApJ...507..794L} {507, 794}

\bibitem[\protect\citeauthoryear{{Lis}, {Menten}, {Serabyn}  \& {Zylka}}{{Lis}
  et~al.}{1994}]{Lis_brick1}
{Lis} D.~C.,  {Menten} K.~M.,  {Serabyn} E.,   {Zylka} R.,  1994, \mn@doi
  [\apjl] {10.1086/187230}, \href
  {http://adsabs.harvard.edu/abs/1994ApJ...423L..39L} {423, L39}

\bibitem[\protect\citeauthoryear{{Longmore}, {Rathborne}, {Bastian}  \& {et
  al}}{{Longmore} et~al.}{2012}]{Brick}
{Longmore} S.~N.,  {Rathborne} J.,  {Bastian} N.,   {et al} 2012, \mn@doi
  [\apj] {10.1088/0004-637X/746/2/117}, \href
  {http://adsabs.harvard.edu/abs/2012ApJ...746..117L} {746, 117}

\bibitem[\protect\citeauthoryear{{Longmore}, {Bally}, {Testi}  \& {et
  al}}{{Longmore} et~al.}{2013a}]{snl_sf}
{Longmore} S.~N.,  {Bally} J.,  {Testi} L.,   {et al} 2013a, \mn@doi [\mnras]
  {10.1093/mnras/sts376}, \href
  {http://adsabs.harvard.edu/abs/2013MNRAS.429..987L} {429, 987}

\bibitem[\protect\citeauthoryear{{Longmore}, {Kruijssen}, {Bally}  \& {et
  al}}{{Longmore} et~al.}{2013b}]{Bricklets}
{Longmore} S.~N.,  {Kruijssen} J.~M.~D.,  {Bally} J.,   {et al} 2013b, \mn@doi
  [\mnras] {10.1093/mnrasl/slt048}, \href
  {http://adsabs.harvard.edu/abs/2013MNRAS.433L..15L} {433, L15}

\bibitem[\protect\citeauthoryear{{Longmore}, {Kruijssen}, {Bastian}  \& {et
  al}}{{Longmore} et~al.}{2014}]{snl_ymc}
{Longmore} S.~N.,  {Kruijssen} J.~M.~D.,  {Bastian} N.,   {et al} 2014, \mn@doi
  [Protostars and Planets VI] {10.2458/azu_uapress_9780816531240-ch013}, \href
  {http://adsabs.harvard.edu/abs/2014prpl.conf..291L} {pp 291--314}

\bibitem[\protect\citeauthoryear{{Longmore} et~al.,}{{Longmore}
  et~al.}{2016}]{Longmore16a}
{Longmore} S.,  et~al., 2016, preprint, \href
  {http://adsabs.harvard.edu/abs/2016arXiv160102654L} {} (\mn@eprint {arXiv}
  {1601.02654})

\bibitem[\protect\citeauthoryear{{Maraston}, {Bastian}, {Saglia}  \& {et
  al}}{{Maraston} et~al.}{2004}]{W3}
{Maraston} C.,  {Bastian} N.,  {Saglia} R.~P.,   {et al} 2004, \mn@doi [\aap]
  {10.1051/0004-6361:20031604}, \href
  {http://adsabs.harvard.edu/abs/2004A%26A...416..467M} {416, 467}

\bibitem[\protect\citeauthoryear{{Matzner} \& {McKee}}{{Matzner} \&
  {McKee}}{2000}]{matzner00}
{Matzner} C.~D.,  {McKee} C.~F.,  2000, \mn@doi [\apj] {10.1086/317785}, \href
  {http://adsabs.harvard.edu/abs/2000ApJ...545..364M} {545, 364}

\bibitem[\protect\citeauthoryear{{McMullin}, {Waters}, {Schiebel}, {Young}  \&
  {Golap}}{{McMullin} et~al.}{2007}]{casa}
{McMullin} J.~P.,  {Waters} B.,  {Schiebel} D.,  {Young} W.,   {Golap} K.,
  2007, in {Shaw} R.~A.,  {Hill} F.,   {Bell} D.~J.,  eds,  Astronomical
  Society of the Pacific Conference Series Vol. 376, Astronomical Data Analysis
  Software and Systems XVI. p.~127

\bibitem[\protect\citeauthoryear{{Molinari}, {Swinyard}, {Bally}  \& {et
  al}}{{Molinari} et~al.}{2010}]{higal}
{Molinari} S.,  {Swinyard} B.,  {Bally} J.,   {et al} 2010, \mn@doi [\pasp]
  {10.1086/651314}, \href {http://adsabs.harvard.edu/abs/2010PASP..122..314M}
  {122, 314}

\bibitem[\protect\citeauthoryear{{Morris} \& {Serabyn}}{{Morris} \&
  {Serabyn}}{1996}]{CMZ}
{Morris} M.,  {Serabyn} E.,  1996, \mn@doi [\araa]
  {10.1146/annurev.astro.34.1.645}, \href
  {http://adsabs.harvard.edu/abs/1996ARA%26A..34..645M} {34, 645}

\bibitem[\protect\citeauthoryear{{Parker}, {Wright}, {Goodwin}  \&
  {Meyer}}{{Parker} et~al.}{2014}]{Tiger14}
{Parker} R.~J.,  {Wright} N.~J.,  {Goodwin} S.~P.,   {Meyer} M.~R.,  2014,
  \mn@doi [\mnras] {10.1093/mnras/stt2231}, \href
  {http://adsabs.harvard.edu/abs/2014MNRAS.438..620P} {438, 620}

\bibitem[\protect\citeauthoryear{{Plummer}}{{Plummer}}{1911}]{Plummer}
{Plummer} H.~C.,  1911, \mn@doi [\mnras] {10.1093/mnras/71.5.460}, \href
  {http://adsabs.harvard.edu/abs/1911MNRAS..71..460P} {71, 460}

\bibitem[\protect\citeauthoryear{{Portegies Zwart}, {Makino}, {McMillan}  \&
  {Hut}}{{Portegies Zwart} et~al.}{2002}]{pz_gc_disruption}
{Portegies Zwart} S.~F.,  {Makino} J.,  {McMillan} S.~L.~W.,   {Hut} P.,  2002,
  \mn@doi [\apj] {10.1086/324141}, \href
  {http://adsabs.harvard.edu/abs/2002ApJ...565..265P} {565, 265}

\bibitem[\protect\citeauthoryear{{Portegies Zwart}, {McMillan}  \&
  {Gieles}}{{Portegies Zwart} et~al.}{2010}]{ymc_port}
{Portegies Zwart} S.~F.,  {McMillan} S.~L.~W.,   {Gieles} M.,  2010, \mn@doi
  [\araa] {10.1146/annurev-astro-081309-130834}, \href
  {http://adsabs.harvard.edu/abs/2010ARA%26A..48..431P} {48, 431}

\bibitem[\protect\citeauthoryear{{Qin}, {Schilke}, {Rolffs}  \& {et al}}{{Qin}
  et~al.}{2011}]{sgrb2}
{Qin} S.-L.,  {Schilke} P.,  {Rolffs} R.,   {et al} 2011, \mn@doi [\aap]
  {10.1051/0004-6361/201116928}, \href
  {http://adsabs.harvard.edu/abs/2011A%26A...530L...9Q} {530, L9}

\bibitem[\protect\citeauthoryear{{Rathborne}, {Longmore}, {Jackson}  \& {et
  al}}{{Rathborne} et~al.}{2014a}]{Brick_jill}
{Rathborne} J.~M.,  {Longmore} S.~N.,  {Jackson} J.~M.,   {et al} 2014a,
  \mn@doi [\apj] {10.1088/0004-637X/786/2/140}, \href
  {http://adsabs.harvard.edu/abs/2014ApJ...786..140R} {786, 140}

\bibitem[\protect\citeauthoryear{{Rathborne}, {Longmore}, {Jackson}  \& {et
  al}}{{Rathborne} et~al.}{2014b}]{Jill_pdf_2014}
{Rathborne} J.~M.,  {Longmore} S.~N.,  {Jackson} J.~M.,   {et al} 2014b,
  \mn@doi [\apjl] {10.1088/2041-8205/795/2/L25}, \href
  {http://adsabs.harvard.edu/abs/2014ApJ...795L..25R} {795, L25}

\bibitem[\protect\citeauthoryear{{Rathborne} et~al.,}{{Rathborne}
  et~al.}{2015}]{rathborne15}
{Rathborne} J.~M.,  et~al., 2015, \mn@doi [\apj] {10.1088/0004-637X/802/2/125},
  \href {http://adsabs.harvard.edu/abs/2015ApJ...802..125R} {802, 125}

\bibitem[\protect\citeauthoryear{{Reid}, {Menten}, {Zheng}  \& {et al}}{{Reid}
  et~al.}{2009}]{distance}
{Reid} M.~J.,  {Menten} K.~M.,  {Zheng} X.~W.,   {et al} 2009, \mn@doi [\apj]
  {10.1088/0004-637X/700/1/137}, \href
  {http://adsabs.harvard.edu/abs/2009ApJ...700..137R} {700, 137}

\bibitem[\protect\citeauthoryear{{Rosolowsky}, {Dunham}, {Ginsburg}  \& {et
  al}}{{Rosolowsky} et~al.}{2010}]{bgps1}
{Rosolowsky} E.,  {Dunham} M.~K.,  {Ginsburg} A.,   {et al} 2010, \mn@doi
  [\apjs] {10.1088/0067-0049/188/1/123}, \href
  {http://adsabs.harvard.edu/abs/2010ApJS..188..123R} {188, 123}

\bibitem[\protect\citeauthoryear{{Sana}, {Momany}, {Gieles}  \& {et al}}{{Sana}
  et~al.}{2010}]{Sana_tr14}
{Sana} H.,  {Momany} Y.,  {Gieles} M.,   {et al} 2010, \mn@doi [\aap]
  {10.1051/0004-6361/200913688}, \href
  {http://adsabs.harvard.edu/abs/2010A%26A...515A..26S} {515, A26}

\bibitem[\protect\citeauthoryear{{Schuller}, {Menten}, {Contreras}  \& {et
  al}}{{Schuller} et~al.}{2009}]{atlasgal_ref}
{Schuller} F.,  {Menten} K.~M.,  {Contreras} Y.,   {et al} 2009, \mn@doi [\aap]
  {10.1051/0004-6361/200811568}, \href
  {http://adsabs.harvard.edu/abs/2009A%26A...504..415S} {504, 415}

\bibitem[\protect\citeauthoryear{{Stolte}, {Hu{\ss}mann}, {Morris}  \& {et
  al}}{{Stolte} et~al.}{2014}]{Stolte_arches}
{Stolte} A.,  {Hu{\ss}mann} B.,  {Morris} M.~R.,   {et al} 2014, \mn@doi [\apj]
  {10.1088/0004-637X/789/2/115}, \href
  {http://adsabs.harvard.edu/abs/2014ApJ...789..115S} {789, 115}

\bibitem[\protect\citeauthoryear{{Urquhart}, {Moore}, {Schuller}  \& {et
  al}}{{Urquhart} et~al.}{2013}]{atlasgal_hii}
{Urquhart} J.~S.,  {Moore} T.~J.~T.,  {Schuller} F.,   {et al} 2013, \mn@doi
  [\mnras] {10.1093/mnras/stt287}, \href
  {http://adsabs.harvard.edu/abs/2013MNRAS.431.1752U} {431, 1752}

\bibitem[\protect\citeauthoryear{{Vesperini}}{{Vesperini}}{2001}]{Vesperini_GC}
{Vesperini} E.,  2001, \mn@doi [\mnras] {10.1046/j.1365-8711.2001.04072.x},
  \href {http://adsabs.harvard.edu/abs/2001MNRAS.322..247V} {322, 247}

\bibitem[\protect\citeauthoryear{{Walker}, {Longmore}, {Bastian}, {Kruijssen},
  {Rathborne}, {Jackson}, {Foster}  \& {Contreras}}{{Walker}
  et~al.}{2015}]{Walker15}
{Walker} D.~L.,  {Longmore} S.~N.,  {Bastian} N.,  {Kruijssen} J.~M.~D.,
  {Rathborne} J.~M.,  {Jackson} J.~M.,  {Foster} J.~B.,   {Contreras} Y.,
  2015, \mn@doi [\mnras] {10.1093/mnras/stv300}, \href
  {http://adsabs.harvard.edu/abs/2015MNRAS.449..715W} {449, 715}

\makeatother
\end{thebibliography}

\bsp

\label{lastpage}

\end{document}